\documentclass[preprint,12pt]{elsarticle}

 \usepackage{graphics}

\usepackage{amssymb}

\journal{Nuclear Physics A}

\begin{document}

\begin{frontmatter}



\title{Role of the mass asymmetry of reaction on the geometry of vanishing flow.}
\author{Supriya Goyal}
\ead{supriyagoyal.pu@gmail.com,ashuphysics@gmail.com}
\address{Department of Physics,
Panjab University, Chandigarh 160 014, India}

\begin{abstract}
We study the transverse flow throughout the mass asymmetry range
as a function of the impact parameter, keeping the total mass of
the system fixed. We find that the geometry of vanishing flow
(GVF) i.e. the impact parameter at which flow vanishes and its
mass dependence is quite sensitive to the mass asymmetry of the
reaction. With increase in the mass asymmetry, the value of GVF
decreases, while its mass dependence increases. Our results
indicate the sizable role of mass asymmetry on GVF as on balance
energy.
\end{abstract}


\end{frontmatter}

\section{Introduction}

Reaction dynamics at low incident energies is mainly governed by
attractive mean field which prompts the emission of particles into
the backward hemisphere, whereas at higher incident energies, due
to the dominance of repulsive binary nucleon-nucleon (nn)
collisions, particle emission takes place into the forward
hemisphere. While going from the low to higher incident energies,
the in-plane flow of particles, also known as collective
transverse flow, disappears at a particular energy. This
particular value of energy is called the Energy of Vanishing Flow
(EVF) or alternatively, the Balance Energy ($E_{bal}$)
\cite{1,2,3,4,5,6,7,8,9,10}. Extensive investigations have been
done to calculate the accurate value of $E_{bal}$ in the last 2-3
decades, both experimentally as well as theoretically
\cite{1,2,3,4,5,6,7,8,9,10}.
\par
The balance energy is found to be very sensitive towards the
nuclear matter equation of state, nn cross-section
\cite{2,3,4,5,6,7,8,9,10}, size of the system \cite{4}, impact
parameter \cite{2,5,7,9,10}, mass asymmetry of the reaction
\cite{11} and incident energy of the projectile \cite{12}. The
mass dependence of balance energy is found to obey the power law
behavior ($\propto A_{TOT}^{\tau}$; where $A_{TOT}$ is the mass of
projectile+mass of target) and ${\tau}$ = -1/3 has been reported
in the literature \cite{4}, whereas recent attempts suggested a
deviation from the above mentioned power law \cite{6,7,8}.
\par
In one of the recent studies, Goyal and Puri \cite{11} found for
the first time, the role of the mass asymmetry of reaction
(defined as $\eta$ = $A_{T}-A_{p}$/$A_{T}+A_{P}$; where $A_{T}$
and $A_{P}$ are the masses of the target and projectile,
respectively) on $E_{bal}$ and its mass dependence. It has been
found that almost independent of the system mass as well as impact
parameter, for large asymmetries (${\eta}=0.7$), the effect of
asymmetry can be 15\% with momentum dependent interactions.
\par
It is well mentioned in the literature that the $E_{bal}$
increases approximately linearly as a function of impact parameter
for symmetric reactions (${\eta}=0$). The linear behavior depends
upon the nuclear equation of state \cite{13}, mass of the system
\cite{13} as well as the mass asymmetry \cite{14}. From the
isolated studies with symmetric systems, it has been found that
linear behavior of $E_{bal}$ with colliding geometry decreases
with increase in system mass \cite{13,14,15}. For the mass
asymmetric systems, the effect of colliding geometries on the
$E_{bal}$ is found for the first time by Goyal \cite{14}. It has
been predicted that the linear dependence of $E_{bal}$ on
colliding geometries increases with increase in mass asymmetry for
each fixed mass of the system.
\par
In a recent study by Puri {\it et al.} \cite{16}, it has been
found for the first time that while going from perfectly central
collisions to most peripheral ones, the collective transverse flow
passes through a maximum (at low values of impact parameter), a
zero value (at some intermediate value of impact parameter), and
achieves negative values (at large values of impact parameter), at
a fixed incident energy. The intermediate value of impact
parameter where collective transverse flow vanishes is termed as
Geometry of Vanishing Flow (GVF) \cite{16}. It has been found that
mass dependence of GVF is insensitive to the nuclear matter
equation of state and momentum dependent interactions, whereas it
is quite sensitive to the binary nn cross-section \cite{16}. It is
noted that the study was done only for the symmetric systems.
Therefore, in the present study we aim to find the role of mass
asymmetry on the GVF at a fixed value of energy. The present
calculations are done with Quantum Molecular Dynamics (QMD) model
\cite{17}, which is found to explain the experimental results of
the mass and impact parameter dependence of the balance energy
(for symmetric systems) very nicely \cite{8}. The model is
explained in Section II. Results and discussion are explained in
Section III and finally we summarizes the results in Section IV.

\section{The quantum molecular dynamics model}

The quantum molecular dynamics model (QMD) is a n-body theory and
simulates the reaction on an event by event basis \cite{17}. The
explicit two and three-body interactions in the model, preserves
the fluctuations and correlations which are important for {\it
N}-body phenomenon such as multifragmentation \cite{17}.
\par
In QMD model, each nucleon ${\it \alpha}$ is represented by a
Gaussian wave packet with a width of $\sqrt{\it L}$ centered
around the mean position {\it $\vec{r}_{\alpha}$}(t) and mean
momentum {\it $\vec{p}_{\alpha}$}(t) \cite{17}:
\begin{equation}
\phi_{\alpha}(\vec{r},\vec{p},t)=\frac{1}{\left(2\pi
L\right)^{3/4}}e^{\left[-\left\{\vec{r}-\vec{r}_{\alpha}(t)\right\}^2/4L\right]}
e^{\left[i\vec{p}_{\alpha}(t)\cdot\vec{r}/\hbar\right]}.
\end{equation}
The Wigner distribution of a system with ${\it A_{T}+A_{P}}$
nucleons is given by
\begin{equation}
f(\vec{r},\vec{p},t)=\sum_{\alpha =
1}^{A_{T}+A_{P}}\frac{1}{\left(\pi \hbar\right)^{3}}
e^{\left[-\left\{\vec{r}-\vec{r}_{\alpha}(t)\right\}^2/2L\right]}
e^{\left[-\left\{\vec{p}-\vec{p}_{\alpha}(t)\right\}^2
2L/\hbar^{2}\right]^{'}},
\end{equation}
with L = 1.08 $fm^{2}$.
\par The center of each Gaussian (in the
coordinate and momentum space) is chosen by the Monte Carlo
procedure. The momentum of nucleons (in each nucleus) is chosen
between zero and local Fermi momentum
[$=\sqrt{2m_{\alpha}V_{\alpha}(\vec{r})}$; $V_{\alpha}(\vec{r})$
is the potential energy of nucleon $\alpha$]. Naturally, one has
to take care that the nuclei, thus generated, have right binding
energy and proper root mean square radii.
\par
The centroid of each wave packet is propagated using the classical
equations of motion \cite{17}:
\begin{equation}
\frac {d\vec{r}_{\alpha}}{dt} = \frac {dH}{d\vec{p}_{\alpha}},
\end{equation}
\begin{equation}
\frac {d\vec{p}_{\alpha}}{dt} = -\frac {dH}{d\vec{r}_{\alpha}},
\end{equation}
where the Hamiltonian is given by
\begin{equation}
H=\sum_{\alpha} \frac {\vec{p}_{\alpha}^{2}}{2m_{\alpha}} + V
^{tot}.
\end{equation}
 Our total interaction potential $V^{tot}$ reads as \cite{17}
\begin{equation}
V^{tot} = V^{Loc} + V^{Yuk} + V^{Coul} + V^{MDI},
\end{equation}
with
\begin{equation}
V^{Loc} = t_{1}\delta(\vec{r}_{\alpha}-\vec{r}_{\beta})+
t_{2}\delta(\vec{r}_{\alpha}-\vec{r}_{\beta})
\delta(\vec{r}_{\alpha}-\vec{r}_{\gamma}),
\end{equation}
\begin{equation}
V^{Yuk}=t_{3}e^{-|\vec{r}_{\alpha}-\vec{r}_{\beta}|/m}/\left(|\vec{r}_{\alpha}-\vec{r}_{\beta}|/m\right),
\end{equation}
with ${\it m}$ = 1.5 fm and $\it{t_{3}}$ = -6.66 MeV.
\par
The static (local) Skyrme interaction can further be parametrized
as:
\begin{equation}
U^{Loc}=\alpha\left(\frac{\rho}{\rho}_o\right)+
\beta\left(\frac{\rho}{\rho}_o\right)^{\gamma}.
\end{equation}
Here $\alpha, \beta$ and $\gamma$ are the parameters that define
equation of state. The momentum dependent interaction is obtained
by parameterizing the momentum dependence of the real part of the
optical potential. The final form of the potential reads as
\begin{equation}
U^{MDI}\approx t_{4}ln^{2}[t_{5}({\it\vec{p}_{\alpha}}-{\it
\vec{p}_{\beta}})^{2}+1]\delta({\it \vec{r}_{\alpha}}-{\it
\vec{r}_{\beta}}).
\end{equation}
Here ${\it t_{4}}$ = 1.57 MeV and ${\it t_{5}}$ = 5 $\times
10^{-4} MeV^{-2}$. A parameterized form of the local plus momentum
dependent interaction (MDI) potential (at zero temperature) is
given by
\begin{equation}
U=\alpha \left({\frac {\rho}{\rho_{0}}}\right) + \beta
\left({\frac {\rho}{\rho_{0}}}\right)+ \delta
ln^{2}[\epsilon(\rho/\rho_{0})^{2/3}+1]\rho/\rho_{0}.
\end{equation}
The parameters $\alpha$, $\beta$, and $\gamma$ in above Eq. (11)
must be readjusted in the presence of momentum dependent
interactions so as to reproduce the ground state properties of the
nuclear matter. The set of parameters corresponding to different
equations of state can be found in Ref. \cite{17}.

\section{Results and discussion}

For the present study, we simulated the reactions of
$^{20}_{10}Ne+^{20}_{10}Ne$ ($\eta = 0$),
$^{17}_{8}O+^{23}_{11}Na$ ($\eta = 0.1$),
$^{14}_{7}N+^{26}_{12}Mg$ ($\eta = 0.3$),
$^{10}_{5}B+^{30}_{14}Si$ ($\eta = 0.5$), $^{6}_{3}Li+^{34}_{16}S$
($\eta = 0.7$), and $^{3}_{2}He+^{37}_{17}Cl$ ($\eta = 0.9$), for
$A_{TOT}$ = 40, $^{40}_{20}Ca+^{40}_{20}Ca$ ($\eta = 0$),
$^{36}_{18}Ar+^{44}_{20}Ca$ ($\eta = 0.1$),
$^{28}_{14}Si+^{52}_{24}Cr$ ($\eta = 0.3$),
$^{20}_{10}Ne+^{60}_{28}Ni$ ($\eta = 0.5$),
$^{10}_{5}B+^{70}_{32}Ge$ ($\eta = 0.7$), and
$^{6}_{3}Li+^{74}_{34}Se$ ($\eta = 0.9$), for $A_{TOT}$ = 80,
$^{80}_{36}Kr+^{80}_{36}Kr$ ($\eta = 0$),
$^{70}_{32}Ge+^{90}_{40}Zr$ ($\eta = 0.1$),
$^{54}_{26}Fe+^{106}_{48}Cd$ ($\eta = 0.3$),
$^{40}_{20}Ca+^{120}_{52}Te$ ($\eta = 0.5$),
$^{24}_{12}Mg+^{136}_{58}Ce$ ($\eta = 0.7$), and
$^{6}_{3}Li+^{154}_{64}Gd$ ($\eta = 0.9$), for $A_{TOT}$ = 160,
and $^{120}_{52}Te+^{120}_{52}Te$ ($\eta = 0$),
$^{108}_{48}Cd+^{132}_{56}Ba$ ($\eta = 0.1$),
$^{84}_{38}Sr+^{156}_{66}Dy$ ($\eta = 0.3$),
$^{60}_{28}Ni+^{180}_{74}W$ ($\eta = 0.5$),
$^{36}_{18}Ar+^{204}_{82}Pb$ ($\eta = 0.7$), and
$^{7}_{3}Li+^{233}_{92}U$ ($\eta = 0.9$), for $A_{TOT}$ = 240 at
full range of colliding geometries ranging from the central to
peripheral collisions in small steps of 0.25 and at a fixed
incident energy of 200 MeV/nucleon. We have varied ${\eta}$ from 0
to 0.9 for every system mass $A_{TOT}$ =40, 80, 160, and 240. Soft
equation of state with momentum dependent interaction (SMD) is
used along with energy dependent Cugnon cross-section.
\par
To calculate the $E_{bal}$, we use {\it directed transverse
momentum $<P^{dir}_{x}>$}, which is defined as:
\begin{equation}
\langle P_{x}^{dir}\rangle=\frac{1}{A}\sum_i {\rm
sign}\{Y(i)\}~{\bf{p}}_{x}(i),
\end{equation}
where $Y(i)$ and ${\bf{p}}_{x}(i)$ are the rapidity distribution
and transverse momentum of $i^{th}$ particle, respectively.
\par
In Fig. 1, we display the $<P^{dir}_{x}>$ as a function of reduced
impact parameter ($b/b_{max}$; where $b_{max}$  = radius of
projectile + radius of target) for ${\eta}$ = 0 - 0.9. The study
is done for different mass ranges. All reactions are followed till
200 fm/c, where $<P^{dir}_{x}>$ saturates. Symbols are explained
in the caption of the figure. Lines are just to guide the eye. As
expected, in all cases i.e. for all values of ${\eta}$ and
$A_{TOT}$, $<P^{dir}_{x}>$ first increases with $b/b_{max}$,
reaches a maximal value and after passing through a zero at some
intermediate value of impact parameter, attains negative values at
large $b/b{max}$. The trend is uniform throughout the mass
asymmetry range i.e. from ${\eta}$ = 0-0.9 for every $A_{TOT}$.
The value of GVF (impact parameter at which $<P^{dir}_{x}>$
attains a zero) varies with ${\eta}$ and $A_{TOT}$. For lighter
systems and larger ${\eta}$, the value of GVF is small compared to
the heavier systems and smaller ${\eta}$.
\par
In Fig. 2, we display GVF as a function of ${\eta}$ for $A_{TOT}$
= 40, 80, 160, and 240. Symbols are explained in the caption of
the figure. Lines are just to guide the eye. The percentage
variation in GVF while going from ${\eta}$ = 0 to 0.9 is -62\%,
-42.03\%, -40.91\%, and -21.21\%, respectively for $A_{TOT}$ = 40,
80, 160, and 240. The negative signs indicates that GVF decreases
with increase in ${\eta}$. It is very clear from the figure that
with increase in system mass, the effect of mass asymmetry of the
reaction on GVF decreases. The effect is similar to as predicted
for $E_{bal}$ \cite{11}. It is well known that with increase in
${\eta}$ and impact parameter, the $E_{bal}$ increases while with
increase in $A_{TOT}$, $E_{bal}$ decreases. This is due to the
decrease in nn collisions with increase in ${\eta}$ and impact
parameter, and increase in Coulomb repulsion with increase in
$A_{TOT}$. The present study has been done at fixed incident lab
energy, therefore, the effective center-of-mass energy also
decreases as ${\eta}$ increases for every fixed system mass. To
compensate all these factors, the value of impact parameter, where
flow vanishes, decreases as ${\eta}$ increases.
\par
In Fig. 3, we display GVF as a function of $A_{TOT}$ for each
${\eta}$. Symbols are explained in the caption of the figure.
Lines are power law fits ($\propto A_{TOT}^{\tau}$). The
percentage variation in GVF while going from $A_{TOT}$ = 40 to 240
is 98\%, 94.12\%, 104.17\%, 121.43\%, 174.19\%, and 310.53\%,
respectively, for ${\eta}$ = 0, 0.1, 0.3, 0.5, 0.7, and 0.9. This
shows that the dependence of GVF on system mass increases with
increase in ${\eta}$. We also display the results of $Hard^{40}$,
$HMD^{40}$, $SMD^{40}$, and $SMD^{cug}$ for symmetric systems with
total mass ranging from 80 to 262 and at an incident energy of 150
MeV/nucleon. Values are taken from Ref. \cite{16}. Superscripts
represents the values of the cross-sections.
\par
From the value of GVF for ${\eta}$ = 0, at 200 MeV/nucleon and 150
MeV/nucleon using $SMD^{cug}$ equation of state, we found that GVF
depends on the incident energy. It was found in Ref. \cite{16}
that mass dependence of GVF is very sensitive to in-medium nn
cross-section. From the figure, it is clear that the difference in
the value of GVF due to the change in cross-section i.e. between
$SMD^{cug}({\eta} = 0)$ and $SMD^{40}({\eta} = 0)$ at 150
MeV/nucleon is same as between $SMD^{cug}({\eta} = 0, E = 150
MeV/nucleon)$ and $SMD^{cug}({\eta} = 0, E = 200 MeV/nucleon)$.
Very interestingly, it shows that for symmetric systems, GVF is
equally sensitive to nn cross-section and incident energy.
\par
In Fig. 4, we display the value of ${\tau}$ as a function of
${\eta}$. Symbols are explained in the caption of the figure. Line
is just to guide the eye. The value of ${\tau}$ increases with
increase in ${\eta}$. For ${\eta}$ = 0, the values of ${\tau}$ at
150 MeV/nucleon are 0.24$\pm$0.01, 0.23$\pm$0.02, 0.25$\pm$0.01,
and 0.44$\pm$0.02, respectively, for $Hard^{40}$, $HMD^{40}$,
$SMD^{40}$, and $SMD^{cug}$ equations of state and are displayed
in the figure \cite{16}. The difference in values of ${\tau}$ at
${\eta}$ = 0 is due to the variation of incident energy and mass
range for which ${\tau}$ is calculated \cite{16}.
\par
In Fig. 5, we display the mean field ($<P^{dir}_{x}>_{mf}$) and
binary nn collision ($<P^{dir}_{x}>_{coll}$) contribution of total
($<P^{dir}_{x}>$) as a function of the colliding geometry. As
expected, ($<P^{dir}_{x}>_{coll}$) is always positive and
($<P^{dir}_{x}>_{mf}$) is always negative for all colliding
geometries. The shaded areas cover the full range of mass in the
present study i.e. from 40 to 240. Upper(lower) boundary of the
shaded area for ${\eta}$ = 0 represent heavier(lighter) mass,
while for ${\eta}$ = 0.9, the upper boundary of
($<P^{dir}_{x}>_{mf}$) represent heavier mass and of
($<P^{dir}_{x}>_{coll}$) represents lighter mass. For different
${\eta}$, the effect of mean field and binary collisions is
different for different masses at and around respective GVF, thus
making GVF and its mass dependence quite sensitive to mass
asymmetry.

\section{Summary}

In short, we have studied the collective transverse flow
throughout the mass asymmetry range ${\eta}$ = 0 - 0.9, as a
function of impact parameter. Study is done for different mass
ranges. We found that the geometry of vanishing flow (GVF) is
quite sensitive to the mass asymmetry. For each ${\eta}$, the mass
dependence of GVF follows a power law behavior and dependence on
the system mass increases with increase in ${\eta}$. The present
study concludes that mass asymmetry has a significant role on the
geometry of vanishing flow.

\section{Acknowledgement}

Author is thankful to Dr. Rajeev K. Puri for interesting and
constructive discussions. This work is supported by a research
grant from the Council of Scientific and Industrial Research
(CSIR), Govt. of India, vide grant No. 09/135(0563)/2009-EMR-1.
\bibliographystyle{elsarticle-num}

\newpage
{\Large \bf Figure Captions}\\

{\bf FIG. 1.} (Color online) $<P^{dir}_{x}>$ (MeV/c) as a function
of reduced impact parameter ($b/b_{max}$) for different system
masses. The results for different mass asymmetries $\eta$ = 0,
0.1, 0.3, 0.5, 0.7, and 0.9 are represented, respectively, by the
solid squares, circles, triangles, inverted triangles, diamonds,
and stars. Results are at an incident energy of 200 MeV/nucleon.\\

{\bf FIG. 2.} (Color online) The geometry of vanishing flow (GVF)
as a function of ${\eta}$ for different system masses. The results
for different system masses ($A_{TOT}$) = 40, 80, 160, and 240 are
represented, respectively, by the half filled squares, circles, triangles, and inverted triangles.\\

{\bf FIG. 3.} (Color online) The geometry of vanishing flow (GVF)
as a function of system mass. The results for different mass
asymmetries $\eta$ = 0, 0.1, 0.3, 0.5, 0.7, and 0.9 are
represented, respectively, by the solid squares, circles,
triangles, inverted triangles, diamonds, and stars. Lines are the
power law fits. The values of GVF at 150 MeV/nucleon and
${\eta}$=0 are represented by open circles, triangles, inverted
triangles, and diamonds, respectively for $Hard^{40}$, $HMD^{40}$,
$SMD^{40}$, and $SMD^{cug}$ equations of state \cite{16}.\\

{\bf FIG. 4.} (Color online) The value of ${\tau}$ as a function
of ${\eta}$. The result of present study is shown by open squares.
Open circles, triangles, inverted triangles, and diamonds,
represents the value of ${\tau}$, respectively for $Hard^{40}$,
$HMD^{40}$, $SMD^{40}$, and $SMD^{cug}$ equations of state at 150 MeV/nucleon \cite{16}.\\

{\bf FIG. 5.} (Color online) The decomposition of $<P^{dir}_{x}>$
into mean and binary collision parts as a function of the reduced
impact parameter for different mass asymmetries (${\eta}$ = 0 and
0.9).
\begin{figure}[!tb]
\centering \vskip -1.8cm \setlength{\abovecaptionskip}{-10cm}
\setlength{\belowcaptionskip}{0.5cm}
\includegraphics[width=13.8cm]{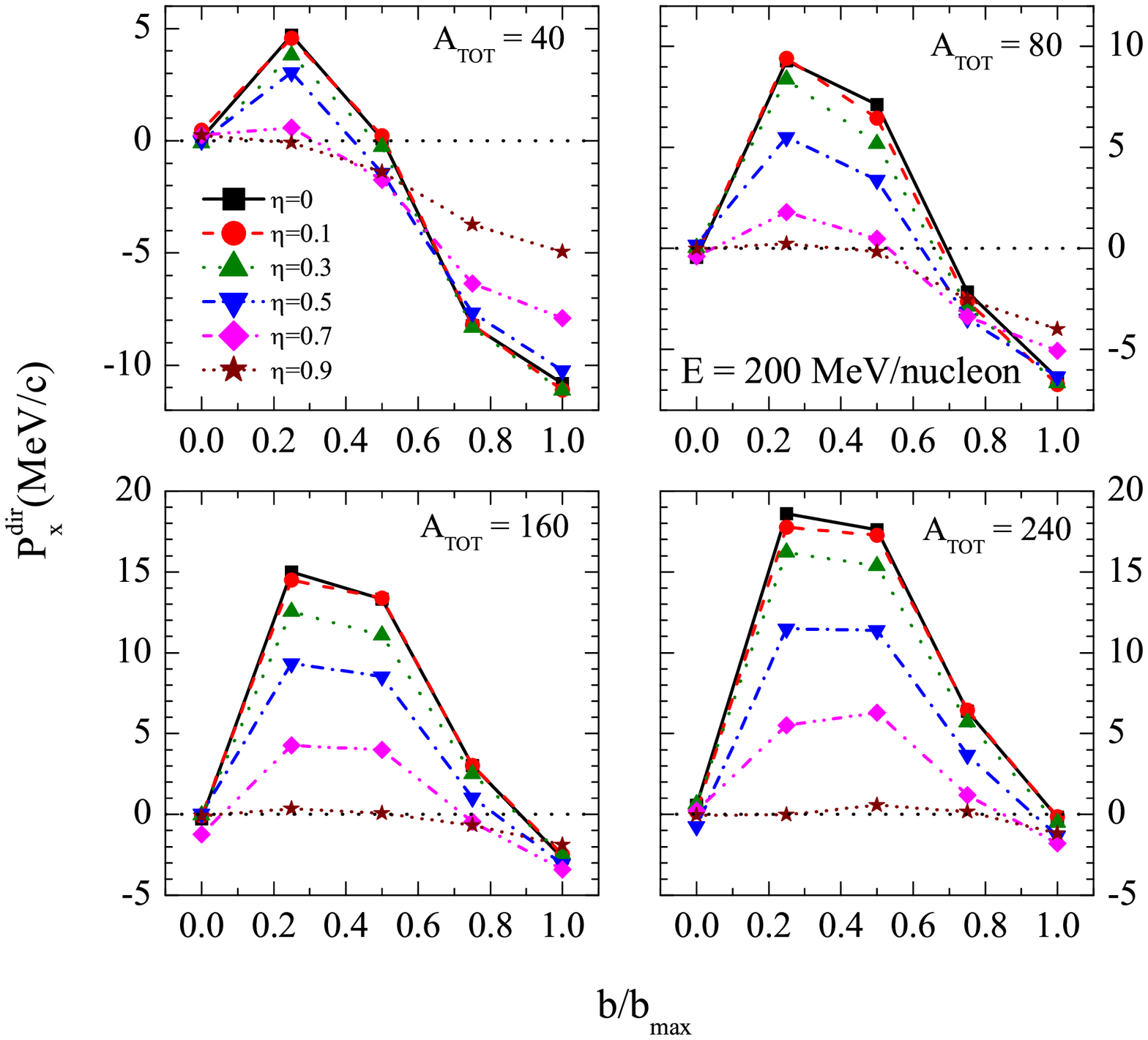}
\end{figure}
\begin{figure}[!tb]
\centering \vskip -1.8cm \setlength{\abovecaptionskip}{-10cm}
\setlength{\belowcaptionskip}{0.5cm}
\includegraphics[width=13.8cm]{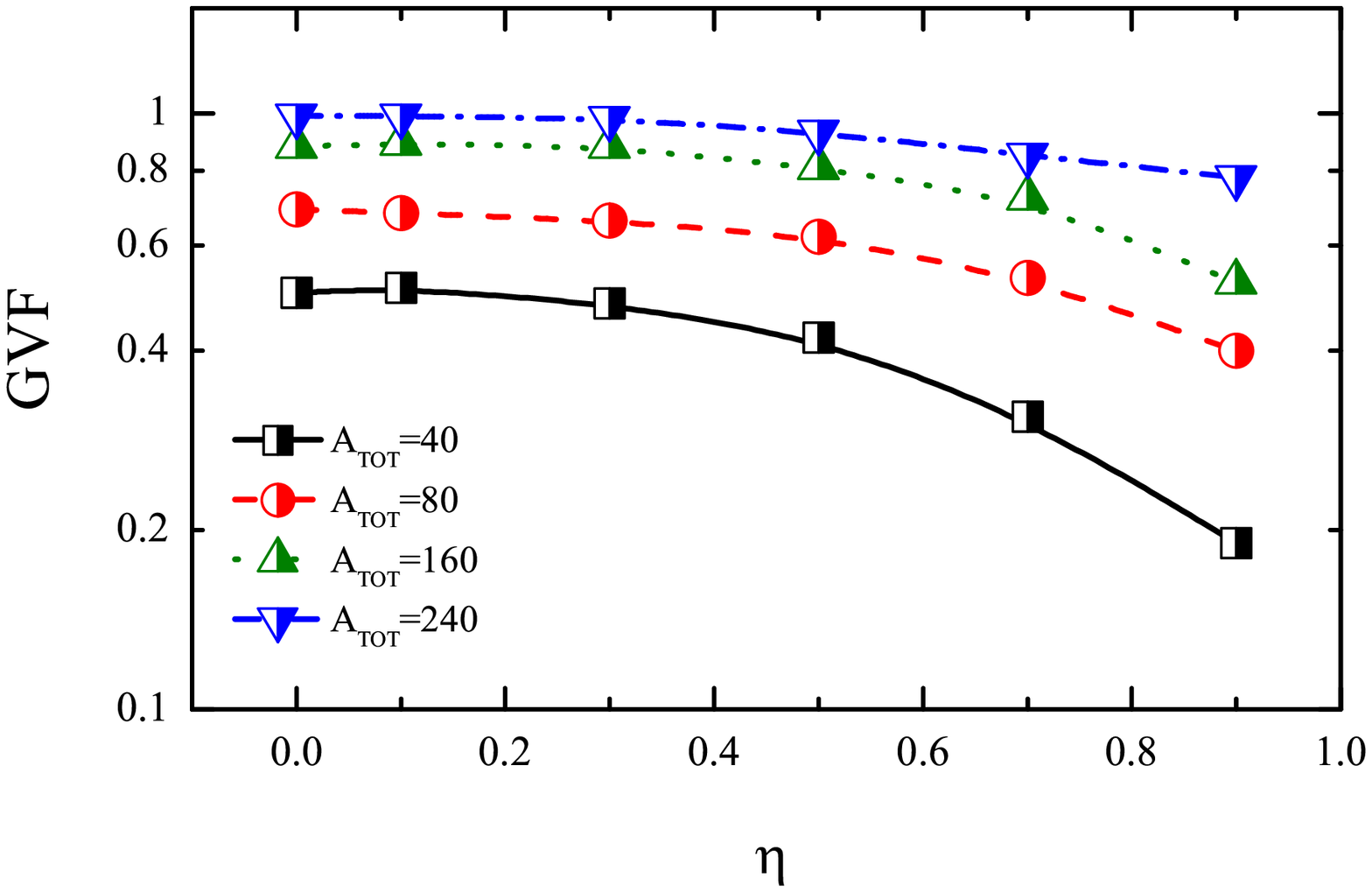}
\end{figure}
\begin{figure}[!tb]
\centering \vskip -1.8cm \setlength{\abovecaptionskip}{-10cm}
\setlength{\belowcaptionskip}{0.5cm}
\includegraphics[width=13.8cm]{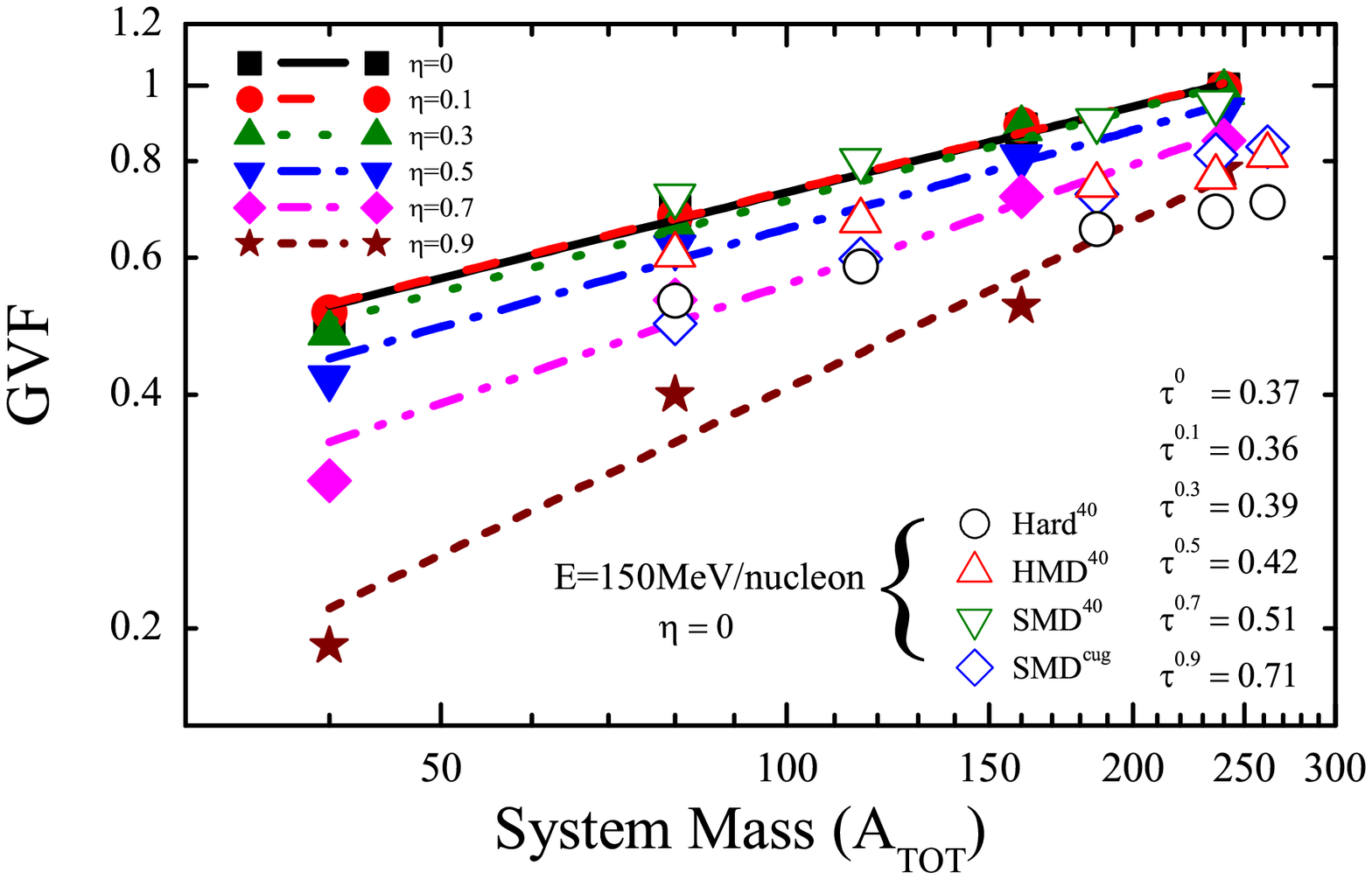}
\end{figure}
\begin{figure}[!tb]
\centering \vskip -1.8cm \setlength{\abovecaptionskip}{-10cm}
\setlength{\belowcaptionskip}{0.5cm}
\includegraphics[width=13.8cm]{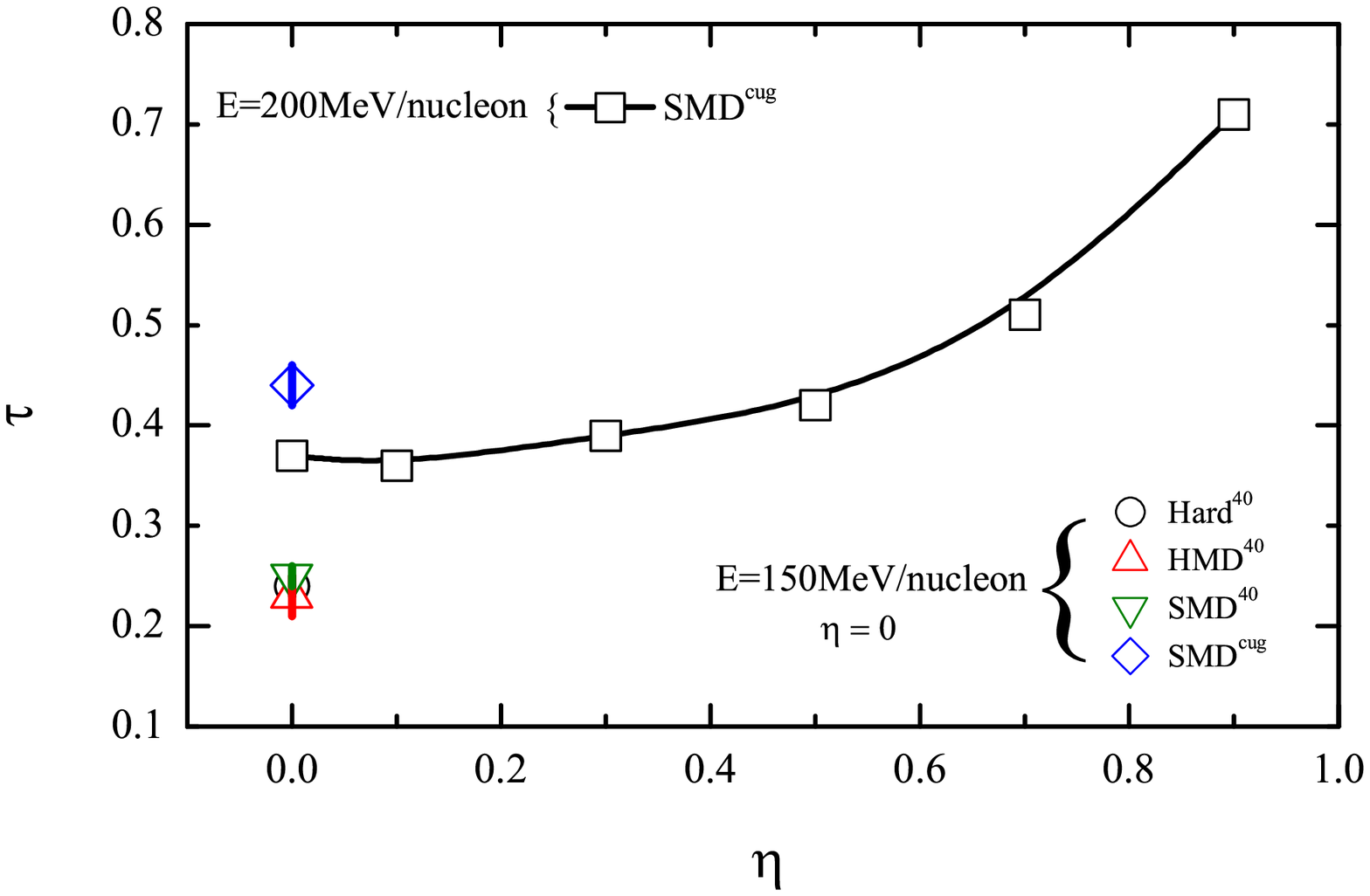}
\end{figure}
\begin{figure}[!tb]
\centering \vskip -1.8cm \setlength{\abovecaptionskip}{-10cm}
\setlength{\belowcaptionskip}{0.5cm}
\includegraphics[width=13.8cm]{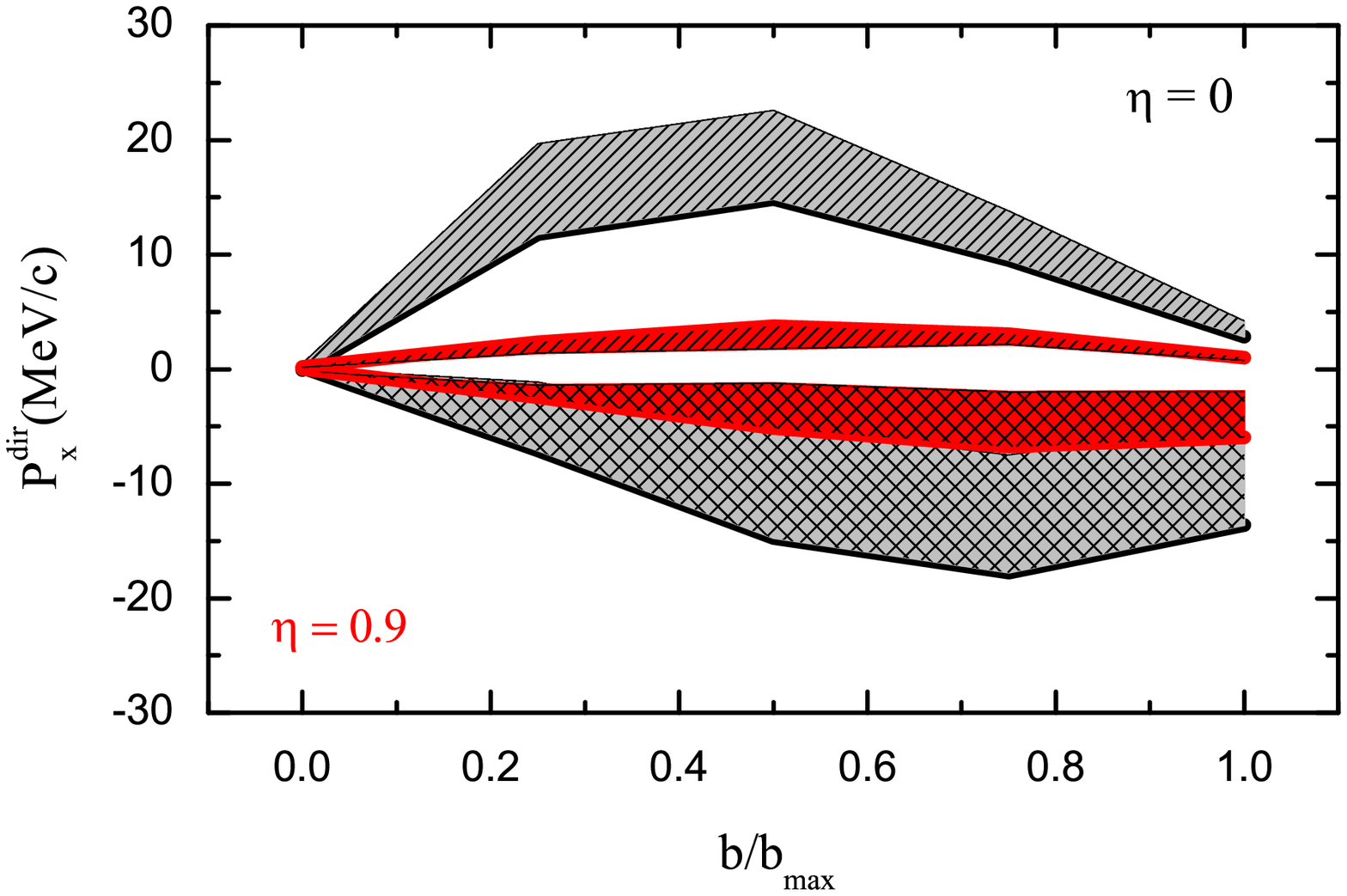}
\end{figure}
\end{document}